\documentclass[aps,pra,nobalancelastpage,twocolumn,superscriptaddress,nofootinbib,amsmath,groupedaddress,longbibliography]{revtex4-1}

\usepackage{CJKutf8}
\usepackage{graphicx}
\usepackage{subfigure}
\usepackage{dcolumn}
\usepackage{bm}
\usepackage{siunitx}
\usepackage{color}
\usepackage{longtable}
\usepackage{amsmath}
\usepackage{amssymb}
\usepackage{array}
\usepackage{multirow}
\usepackage{tablefootnote}
\usepackage{upgreek}
\usepackage{braket}
\usepackage{makecell}
\usepackage[letterpaper,total={7in,9.5in},top=0.75in,left=0.75in]{geometry}
\usepackage[colorlinks=true,linkcolor=blue,citecolor=blue,urlcolor=blue,filecolor=blue]{hyperref}
\usepackage{lipsum}  
\usepackage{xcolor}
\usepackage{ulem} 
\usepackage{orcidlink}

\definecolor{lime}{HTML}{A6CE39}
\DeclareRobustCommand{\orcidicon}{
	\begin{tikzpicture}
	\draw[lime, fill=lime] (0,0) 
	circle [radius=0.16] 
	node[white] {{\fontfamily{qag}\selectfont \tiny ID}};
	\draw[white, fill=white] (-0.0625,0.095) 
	circle [radius=0.007];
	\end{tikzpicture}
	\hspace{-0.3mm}
}

\foreach \x in {A, ..., Z}{\expandafter\xdef\csname orcid\x\endcsname{\noexpand\href{https://orcid.org/\csname orcidauthor\x\endcsname}
			{\noexpand\orcidicon}}
}

\DeclareSIUnit\gauss{G}

\newcolumntype{L}[1]{>{\raggedright\let\newline\\\arraybackslash\hspace{0pt}}m{#1}}
\newcolumntype{C}[1]{>{\centering\let\newline\\\arraybackslash\hspace{0pt}}m{#1}}
\newcolumntype{R}[1]{>{\raggedleft\let\newline\\\arraybackslash\hspace{0pt}}m{#1}}

\newcommand{\IRTransition}{${^3\mathrm{P}_0} - {^3\mathrm{D}_1}$ }
\newcommand{\clocktransition}{${^1\mathrm{S}_0} - {^3\mathrm{P}_0}$ }
\newcommand{\clockstate}{${^3\mathrm{P}_{0}}$ }
\newcommand{\groundstate}{${^1\mathrm{S}_0}$ }

\newcommand{\GreenMotTransition}{${^1\mathrm{S}_0} - {^3\mathrm{P}_1}$ }

\newcommand{\Erecoil}{$\textrm{E}_{\textrm{r}}$}
\newcommand{\newtext}[1]{\textcolor{black}{#1}}
\newcommand{\finalEditnewtext}[1]{\textcolor{black}{#1}}

 \newcommand{\beginsupplement}{%
        \setcounter{table}{0}
        \renewcommand{\thetable}{S\arabic{table}}%
        \setcounter{figure}{0}
        \renewcommand{\thefigure}{S\arabic{figure}}%
     }

\let\oldchi\chi
\renewcommand{\chi}{%
  \raisebox{0.44ex}{$\oldchi$}%
}

\begin{document}
\title{Clock-line-mediated Sisyphus Cooling}

\author{Chun-Chia Chen (陳俊嘉)\,$\orcidA{}$,\textsuperscript{1,2,*,\dag}, Jacob L. Siegel\,$\orcidB{}$,\textsuperscript{1,2,*}, Benjamin D. Hunt\,$\orcidC{}$,\textsuperscript{1,2}, Tanner Grogan\textsuperscript{1,2}, Youssef S. Hassan\,$\orcidD{}$,\textsuperscript{1,2}, Kyle Beloy\,$\orcidE{}$,\textsuperscript{1}, Kurt Gibble\,$\orcidF{}$,\textsuperscript{1,3},  Roger C. Brown\,$\orcidG{}$,\textsuperscript{1}, Andrew D. Ludlow \textsuperscript{1,2,\dag}}

\affiliation{\textsuperscript{1}National Institute of Standards and Technology, 325 Broadway, Boulder, Colorado 80305, USA}
\affiliation{\textsuperscript{2}University of Colorado, Department of Physics, Boulder, Colorado 80309, USA}
\affiliation{\textsuperscript{3}Department of Physics, The Pennsylvania State University, University Park, Pennsylvania 16802, USA}
\date{\today}

\begin{abstract}
We demonstrate sub-recoil Sisyphus cooling using the long-lived $^{3}\mathrm{P}_{0}$ clock state in alkaline-earth-like ytterbium. 
A 1388\,-nm optical standing wave nearly resonant with the $^{3}\textrm{P}_{0}$$\,\rightarrow$$\,^{3}\textrm{D}_{1}$ transition creates a spatially periodic light shift of the $^{3}\textrm{P}_{0}$ clock state. \newtext{Following excitation on the ultranarrow clock transition, we observe Sisyphus cooling in this potential, as the light shift is correlated with excitation to $^{3}\textrm{D}_{1}$ and subsequent spontaneous decay to the $^{1}\textrm{S}_{0}$ ground state.} We observe that cooling enhances the loading efficiency of atoms into a 759\,-nm magic-wavelength \newtext{one-dimensional (1D)} optical lattice, as compared to standard Doppler cooling on the $^{1}\textrm{S}_{0}$$\,\rightarrow\,$$^{3}\textrm{P}_{1}$ transition. 
 Sisyphus cooling yields temperatures below 200\,nK in the weakly confined, transverse dimensions of the 1D optical lattice. 
These lower temperatures improve optical lattice clocks by facilitating the use of shallow lattices with reduced light shifts, while retaining large atom numbers to reduce the quantum projection noise. This Sisyphus cooling can be pulsed or continuous and is applicable to a range of quantum metrology applications. 

\end{abstract}

\begin{CJK*}{UTF8}{min}
\maketitle
\end{CJK*}

\footnote{\label{a}$^*$ These authors contributed equally}
\footnote{\label{b}$^{\dag}$ chenchunchia@gmail.com, andrew.ludlow@nist.gov}

Laser cooling the motion of atoms and molecules enables exquisite quantum control of both their internal and external degrees of freedom. Their rich multilevel structure can be a powerful platform for quantum sensors~\cite{Schmidt2015RMPLudlow}, quantum simulation~\citep{Bloch2012QSimUltracold}, information encoding through quantum control~\cite{Saffman2022Nature, Lukin2022CoherentTransportQC, Ringbauer2022NPhysQudit}, as well as precision measurements for tests of fundamental physics~\cite{Safronova2018RMP}.

As the motivation to exploit more diverse quantum systems has grown, so have the efforts to cool these species. For example, laser cooling of antihydrogen has significantly improved the spectroscopy of antimatter, advancing tests of charge–parity–time invariance~\cite{Baker2021LaserCoolingAntiH}. Remarkable progress has also been made in the field of molecule laser cooling, specifically in the efficient cooling of molecules with nearly-diagonal Franck-Condon factors~\cite{Rosa2004_DiagonalFCF} and those possessing optical cycling centers~\cite{Hudson2022NatureChemistry_OCC, Dickerson2021PRL_OpticalCyclingCenter}. 
\newtext{Beyond cooling more species, the drive for improved quantum control has also motivated deeper cooling techniques. For example, the finite temperature of optical-tweezer-trapped atoms causes random Doppler shifts and position fluctuations, thereby imposing additional constraints on entanglement generation and lifetime~\cite{Lukin2022CoherentTransportQC}. For optical lattice clocks that utilize alkaline-earth(-like) elements, extremely low temperatures can facilitate shallow lattices, which is key to realizing fractional frequency uncertainty at $10^{-18}$ or below~\cite{Katori2015LightshiftStrategy, Brown2017_Hyperpolarizability, Kim2023PRLSrLightshift}. Furthermore, the metastable states in these elements are not just well-suited for next-generation optical atomic clocks, but also for long-lived entangled qubits~\cite{Madjarov2020SrRydbergQubit, Schine2022LonglivedBellstates} and atom interferometry~\cite{Hogan2013_ClockAtomInterferometry}. In these cases, the combination of a strong cycling transition and weaker intercombination transition are ideally used to reach atomic temperatures approaching the $\mu$K level, though deeper cooling is advantageous. This is especially true for alkaline-earth\newtext{(-like)} atoms where the intercombination linewidth is either too narrow or too broad to usefully reach $\mu$K temperatures, with Mg (36\,Hz)~\cite{Rasel2007_quenchcoolingMg} and Hg (1.3\,MHz)~\cite{Katori2008_MercuryLattticeclock} being extreme examples of this.}

\begin{figure}[tb]
\includegraphics[width=0.98\columnwidth]{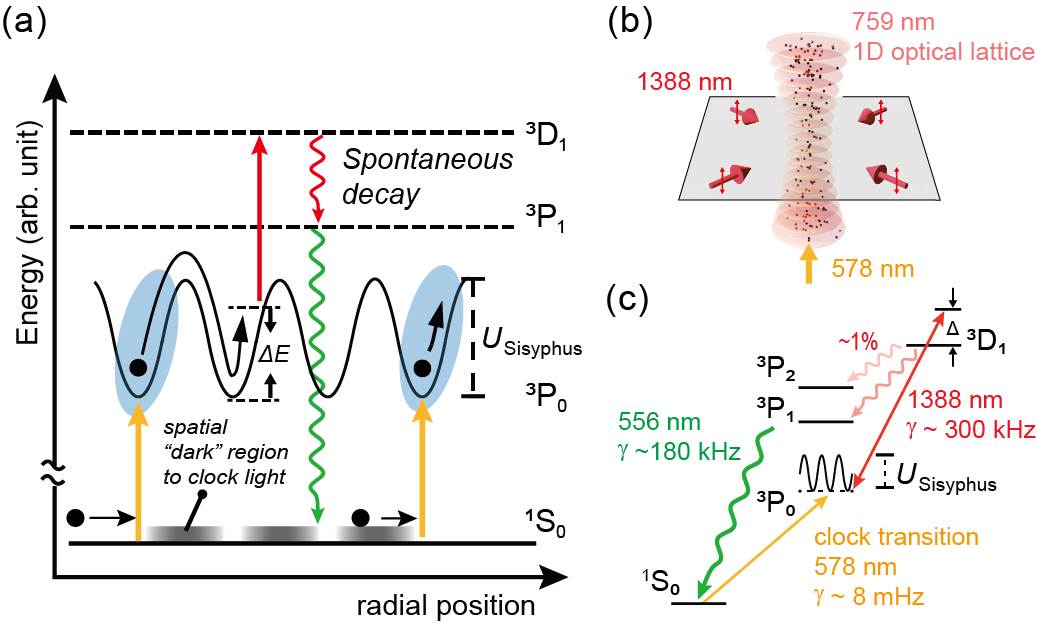}\caption{\label{fig:Setup_scheme} \textbf{Clock-line mediated Sisyphus cooling.} (a) Laser light resonant with the $^1\mathrm{S}_{0}$$\rightarrow$$^3\mathrm{P}_{0}$ clock transition excites atoms near the minima of the spatially varying energy of the $^3\mathrm{P}_{0}$ state.
Atoms then traverse the Sisyphus potential with a depth of $U_\mathrm{Sisyphus}$ before absorbing a $^{3}\mathrm{P}_{0}$$\rightarrow$$^{3}\textrm{D}_{1}$ photon. 
Excitation to the $^{3}\mathrm{D}_{1}$ state is followed by spontaneous decay to the ground state, by way of the $^3\mathrm{P}_1$ state, completing a Sisyphus cycle with a $\Delta E$ energy dissipation.
(b) Two pairs of counter-propagating 1388-nm beams cross orthogonally and form a 2D standing wave in the transverse plane of the 1D optical lattice. The 578-nm clock excitation laser is aligned along the longitudinal axis of the 1D 759-nm magic-wavelength lattice. (c) Simplified Yb level diagram and transitions used in the Sisyphus cooling.
A quasi-closed Sisyphus cooling cycle is established with $\sim 1\%$ branching to the $^{3}\mathrm{P}_{2}$ state.
}
\end{figure}

Here, we adapt a cooling technique proposed for (anti)hydrogen~\cite{Wu2011_SisyphusCoolHydrogen, Chen2019SOLD} to the alkaline-earth-like species ytterbium. The cooling uses an excited-state Sisyphus potential created with a spatially-varying ac-Stark shifting-laser \finalEditnewtext{blue-detuned to the}  $^{3}\textrm{P}_{0}\rightarrow\,^{3}\textrm{D}_{1}$ transition. 
To cool, an ultra-narrow clock laser excites atoms from the \newtext{virtually} unperturbed $^{1}\textrm{S}_{0}$ ground state to the bottom of the $^{3}\textrm{P}_{0}$ Sisyphus potential, see Fig.~\ref{fig:Setup_scheme}(a).
\finalEditnewtext{After atoms lose kinetic energy by climbing the Sisyphus potential, they preferentially absorb $^{3}\textrm{P}_{0}\rightarrow\,^{3}\textrm{D}_{1}$ photons at places away from the potential's minimum due to the higher intensity of the blue-detuned ac-Stark shifting laser.}
Atoms then spontaneously decay to the $^{1}\mathrm{S}_{0}$ state, completing a cooling cycle.
By adjusting the \finalEditnewtext{Rabi frequency} of the clock excitation, we can optimize the technique to achieve either faster cooling or lower temperatures.

Tuning the Stark-shifting laser closer to resonance to increase the Sisyphus potential depth, we cool the atomic sample prepared from the \GreenMotTransition narrow-line magneto-optical trap (MOT). \newtext{Doing so enhances} the number of cold atoms that can be loaded into a 1D magic-wavelength optical lattice~\cite{Katori2003magicwavelength} by a factor of 4.
By then implementing \newtext{1D} Sisyphus cooling on atoms trapped in the lattice, we achieve cooling temperatures below 200\,nK along a weakly \newtext{confined} axis of the lattice.
Instead, with a 2D Sisyphus \newtext{potential} in the \newtext{radial} plane of the 1D optical lattice, we demonstrate 3D cooling, achieving both recoil-limited temperatures along the radial axes as well as cooling along the lattice axis.
We further combine Sisyphus cooling with \finalEditnewtext{energy selective} excitation on the clock transition tuned to a lattice motional sideband. This technique, similar to cyclic cooling~\cite{Pritchard1983CyclicCooling}, reduces the trapped atomic sample's energy distribution along the weak confinement axis. 
Because the Sisyphus cooling scheme utilizes the narrowband clock line shared by alkaline-earth(-like) systems, the technique may readily be applied to elements such as Cd, Hg, Mg, and Ra, which have otherwise less favorable Doppler cooling properties~\cite{Katori2008_MercuryLattticeclock, Katori2019_CdMagiceWavelength, Heckel2016_HgEDM, ANL2007Radium}.

\label{Sec:ExperimentSetup}
Our apparatus and the experimental details to prepare ultracold Yb loaded into a 1D 759\,-nm magic wavelength lattice have been described elsewhere~\cite{McGrew2018_cmlevelAtomicClockresolution}. 
Here, we take laser cooling on the 556\,-nm \GreenMotTransition intercombination transition as our initial condition. We introduce two pairs of counter-propagating 1388\,-nm beams that cross orthogonally and are located in the plane perpendicular to the magic-wavelength lattice axis, see Fig.~\ref{fig:Setup_scheme}(b). The 1388-nm laser beams are collimated, with $\,\sim 1\,\mathrm{mm}$ $1/e^{2}$ diameter. To create a spatially varying ac-Stark shift on the $^{3}\mathrm{P}_{0}$ state, the 1388\,-nm laser is blue-detuned from the \IRTransition transition, as shown in Fig.~\ref{fig:Setup_scheme}(c). The counter-propagating 1388\,-nm laser can be configured either to form an intensity lattice (lin~$\parallel$~lin) or a polarization gradient lattice (lin~$\perp$~lin). We have opted for the lin~$\parallel$~lin configuration, where the minimum light shift corresponds to zero intensity of the ac-Stark shift beam~\cite{SupplInf}, though both configurations yield useful cooling.
We note that multiple dipole allowed transitions that connect the 
$^{3}\mathrm{P}_{0}$ state to higher-lying excited states could be used for implementing the Sisyphus cooling, including $^{3}\mathrm{S}_{1}$ and $^{3}\mathrm{D}_{1}$ states\textcolor{black}{~\cite{SupplInf, Repumpfootnote}}. In our experiment, we choose the $^{3}\mathrm{D}_{1} (\newtext{\textrm{F}=3/2})$ state since the 1388\,-nm laser is already used for detection of the $6s6p\,{^3\mathrm{P}_0}$ state \finalEditnewtext{and have experimentally verified Sisyphus cooling is also effective with the other hyperfine components $^{3}\mathrm{D}_{1} (\newtext{\textrm{F}=1/2})$.}

\begin{figure}[tb]
\centering
\includegraphics[width=0.98\columnwidth]{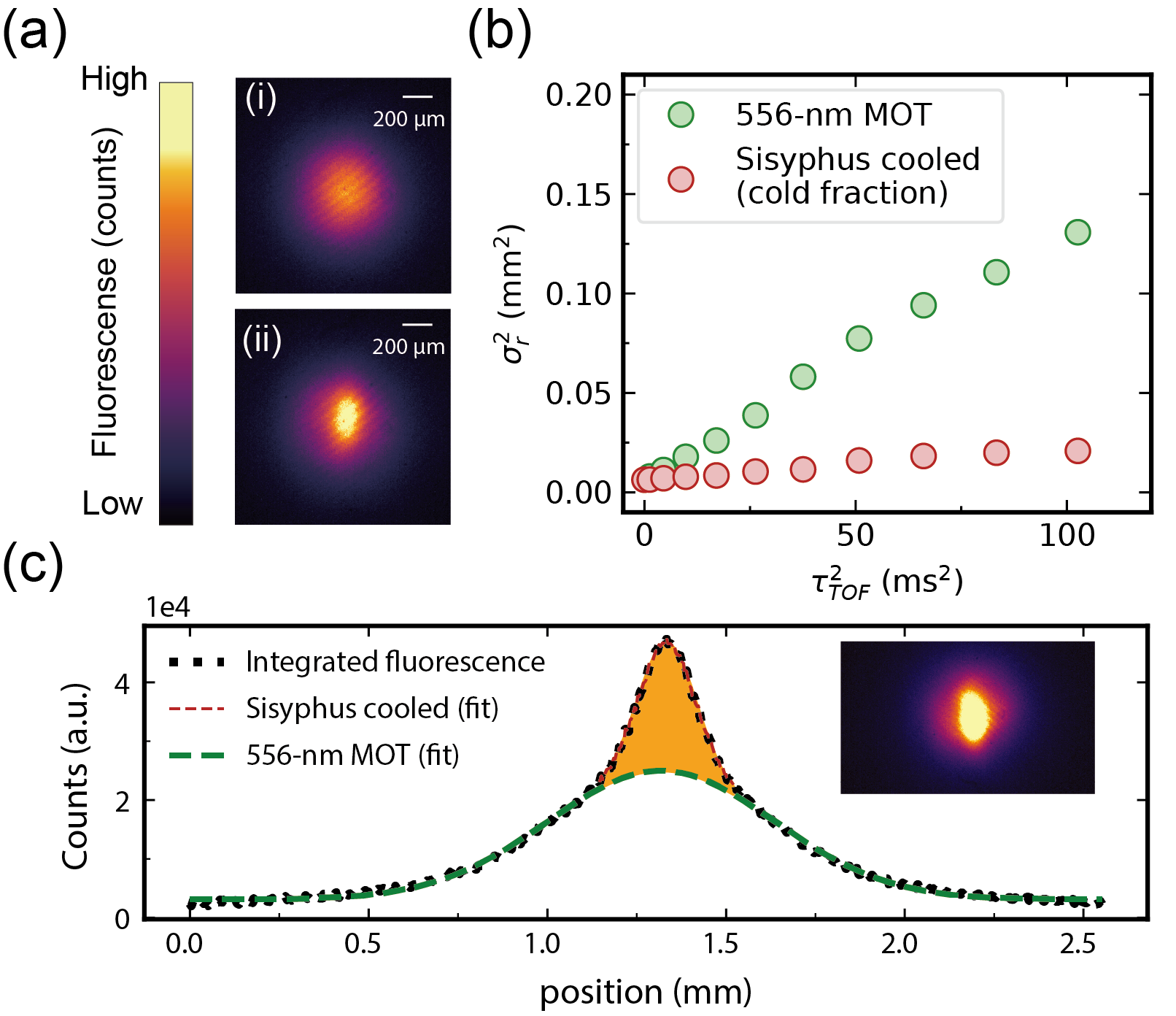}
\caption{\label{fig:freespace_Sisyphuscooling} \textbf{Sisyphus cooling in free space.} (a) Fluorescence images of the atomic sample during free expansion: (i) without and (ii) with Sisyphus cooling applied. (b) TOF trace of the 1$\sigma$ radius obtained from the Gaussian fits to the horizontal, Sisyphus-cooling direction.
The temperature is $\sim$20$\,\mu$K for the narrow-line MOT and $\sim$3$\,\mu$K for the Sisyphus cooled fraction. (c) Integrated fluorescence profiles of the atomic sample showing Sisyphus cooling. Black points are the measured integrated fluorescence along the vertical axis. The shaded orange area and red dashed fit indicates the narrower Gaussian of the Sisyphus cooled fraction\,($\sim$20\%). \newtext{The green dashed fit line shows the broader Gaussian corresponding to the non-Sisyphus-cooled sample from the narrow-line MOT.  }}
\end{figure}

\textit{Sisyphus cooling in free space.---}We first demonstrate Sisyphus cooling \newtext{with untrapped atoms} in free space.
Atoms are cooled in a 556\,-nm narrow-line MOT with a measured temperature of $\sim$\,$20\,\mu$K. \finalEditnewtext{After turning off the 556\,-nm laser, we turn on both the 578\,-nm clock light, characterized by a Rabi frequency exceeding kHz, and the 1388-nm laser for 7\,ms of Sisyphus cooling as the atoms also begin to fall under gravity before taking fluorescence images, see Fig.~\ref{fig:freespace_Sisyphuscooling}(a).} \newtext{The 578\,-nm clock laser needs to be spectrally-narrow and frequency-stable only compared to the desired excitation Rabi rate and not the clock-transition natural linewidth.}
We measure a time-of-flight (TOF) temperature of $\sim$3\,$\mu$K for the Sisyphus-cooled fraction, see Fig.~\ref{fig:freespace_Sisyphuscooling}(b), which is about 20$\%$ of the total atomic sample, see Fig.~\ref{fig:freespace_Sisyphuscooling}(c). Cooling is optimized by blue-detuning the 1388-nm laser to $\Delta\simeq\,+50\,\mathrm{MHz}$ from the \IRTransition~$(\newtext{\mathrm{F}=3/2})$ transition.
This corresponds to a Sisyphus potential depth of $U/h >50$\,kHz ($U/k_{B} >2.4\,\mu$K) in the $^{3}\mathrm{P}_{0}$ state~\cite{SupplInf}.

Adding Sisyphus cooling during the last 10\,ms of the narrow-line MOT and for a further 10\,ms after the MOT is extinguished, we observe enhanced loading into the magic-wavelength optical lattice by a factor of $\sim\,$4. This enhancement works consistently for a broad range of optical lattice trap depths, from $10\,\mathrm{E}_{\mathrm{r}}$ to $100\,\mathrm{E}_{\mathrm{r}}$. Here $\textrm{E}_{\textrm{r}}$ is the lattice photon recoil energy, given by $\textrm{E}_{\textrm{r}} = h^{2}/(2m\lambda_{\textrm{latt}}^{2}
$), where $h$ is Planck’s constant, $m$ is the mass of $^{171}$Yb, and $\lambda_{\textrm{latt}}$ is the 759-nm lattice wavelength. We expect the above-mentioned enhancement can be improved with higher clock-light Rabi frequency, better spatial overlap of the 578-nm clock excitation beam with the green-MOT, and a deeper Sisyphus potential.  

\begin{figure}[tb]
\includegraphics[width=1\columnwidth]{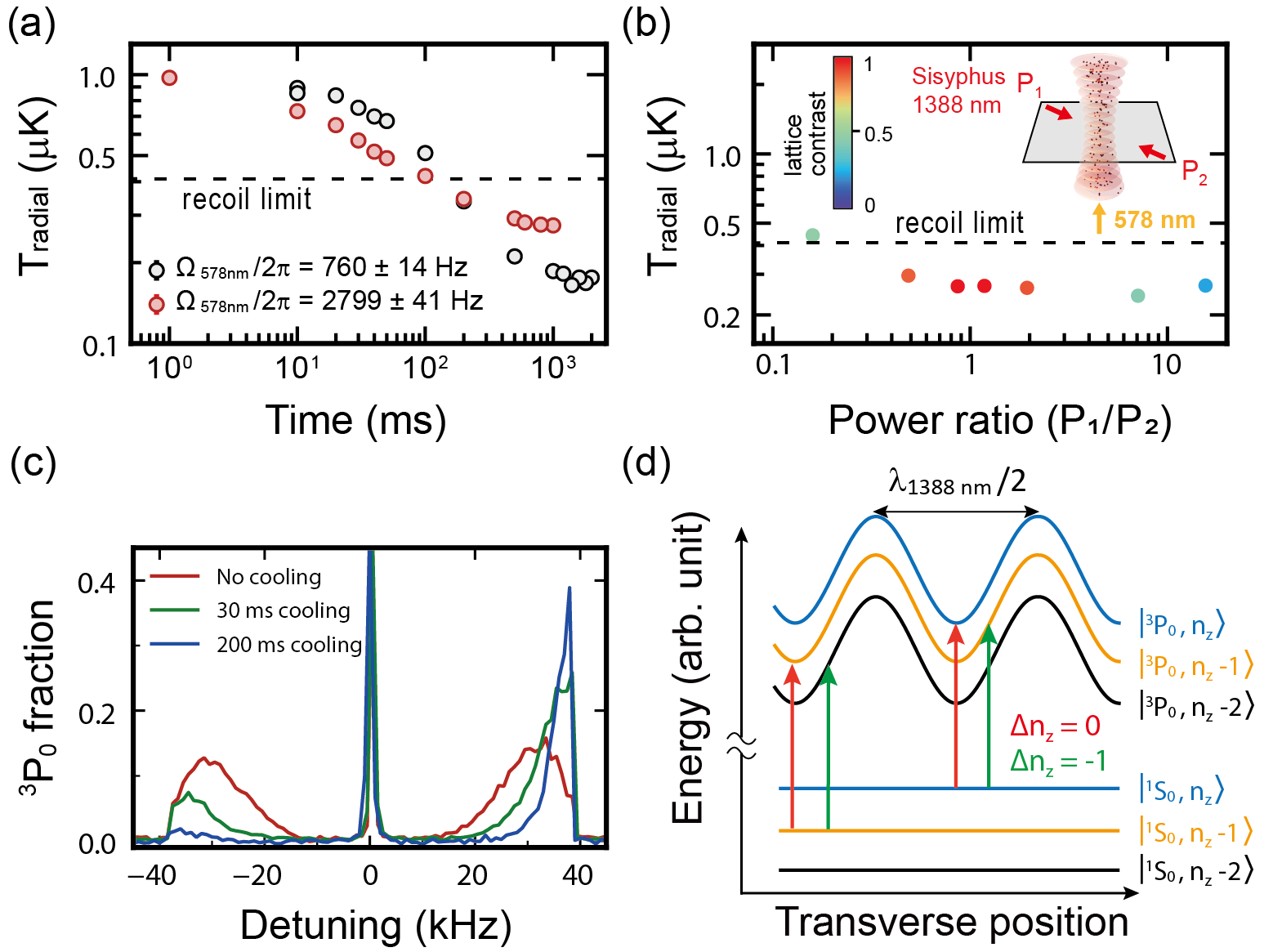}
\caption{\label{fig3:Sisyphus_cooling} \textbf{Sisyphus cooling in a \finalEditnewtext{magic-wavelength} lattice.} (a) Measured radial temperature as a function of 1D Sisyphus cooling time at 60\,$\textrm{E}_{\textrm{r}}$ with different 578-nm clock-excitation Rabi frequencies. (b) \finalEditnewtext{1D} Sisyphus cooling as a function of the power ratio of counter-propagating 1388-nm Sisyphus lattice laser beams for a cooling time of 500\,ms with clock Rabi frequency of 2.8\,kHz\newtext{, where the total 1388-nm laser power remains constant}. The lattice modulation of the ${^3\mathrm{P}_0}$ energy only decreases by at most 27\,\% for a power imbalance ratio of up to 10. (c) Longitudinal sideband spectra for no 2D Sisyphus cooling (red), 30\,ms (green) and 200\,ms (blue) 2D radial Sisyphus cooling, demonstrating an effective 3D cooling. (d) The spatial variation of the ac-Stark shift allows red sideband cooling by moving the lower motional state to resonance with the clock light, which is originally set to drive the carrier transition.}
\end{figure}

\textit{1D sub-recoil Sisyphus cooling in a magic-wavelength lattice.---}
We next demonstrate Sisyphus cooling for atoms already trapped in the 1D lattice. We note that the lattice is magic (light-shift-free) for the \clocktransition clock transition, used here for the initial excitation in the cooling process, see Fig.~\ref{fig3:Sisyphus_cooling}.
We applied Sisyphus cooling along a single radial direction of the 1D lattice with the 1388\,nm frequency set blue-detuned $\Delta \simeq$ 150\,MHz from the $^{3}\textrm{P}_{0}$$\rightarrow$$\,^{3}\textrm{D}_{1}(\newtext{\textrm{F}=3/2})$ transition~\cite{SupplInf}, and then measured the temperature using Doppler spectroscopy. 

Atoms are \newtext{also confined in the magic-wavelength} lattice with a depth of 60\,\Erecoil. The resulting 1D temperatures are shown in Fig.~\ref{fig3:Sisyphus_cooling}(a) as a function of cooling time for two different excitation rates on the 578\,-nm clock transition. \newtext{We compare cooling performance between excitation at two different Rabi frequencies on the clock transition. Stronger clock excitation ($\Omega_{\textrm{578nm}}/2\pi=2.7\,\mathrm{kHz}$) resulted in faster cooling, in this case corresponding to an exponential time constant of $\sim$36(5)\,ms. Because excitation on the clock transition can be the most time-consuming step in the cooling cycle, faster excitation naturally affords faster cooling. However, the higher Rabi frequency also resulted in a higher steady-state cooled temperature of T$\sim$300\,nK. On the other hand, weaker excitation ($\Omega_{\textrm{578nm}}/2\pi=0.760(14)\,\mathrm{kHz}$) gave a lower temperature of T$\sim$165(5)\,nK but with a longer cooling time constant of 123(6)\,ms.} \newtext{Note that the final temperature in these two clock excitation conditions are both} well below the recoil temperature (410\,nK) for the cascaded spontaneous decay ${^3\mathrm{D}_1}-{^3\mathrm{P}_1}-{^1\mathrm{S}_0}$ of the cooling cycle, $k_BT_r =\hbar^2 (k_{D-P}^2 + k_{P-S}^2)/2m$, where $k_{D-P}$($k_{P-S}$) is the wavenumber corresponding to the $^{3}\textrm{D}_{1}$$\rightarrow$$\,^{3}\textrm{P}_{1}$ ($^{3}\textrm{P}_{1}$$\rightarrow$$\,^{1}\textrm{S}_{0}$) transition, and $k_B$ is the Boltzmann constant.

Our interpretation for the sub-recoil temperature and the \finalEditnewtext{decrease of the steady\newtext{-}state temperature with the Rabi frequency} ($\Omega_{\textrm{clock}}$) is as follows. The periodic ac-Stark shift (Sisyphus potential) leads to a spatially dependent excitation profile of the clock light.
\finalEditnewtext{With a deep Sisyphus potential ($U/h \sim$ tens of kHz)~\cite{SupplInf}, atomic motion is quantized.} Away from the Sisyphus potential minimum, ground state atoms are not resonant with the clock excitation due the ac-Stark shift of the \clockstate \,state and therefore are decoupled from the atom-laser interaction. This creates a spatial dark region, see Fig.~\ref{fig:Setup_scheme}(a), where atoms experience periods of darkness to the clock light, an effective ``dark state"~\cite{Zoller1994PRL_LaserCoolingToBEC,Dum1996PRA_Darkstate, Morigi1998PRA_Spatialdark}.
Increasing $\Omega_{\textrm{clock}}$ effectively increases an atom's chances of exiting the dark state. We observe that the sub-recoil cooling performance is robust against power imbalance in the counter-propagating Sisyphus lattice beams. As Fig.~\ref{fig3:Sisyphus_cooling}(b) shows, similar temperatures are obtained with relatively large power imbalances.

\textit{3D cooling in a magic-wavelength lattice.---} We apply Sisyphus cooling in the plane \newtext{transverse to the magic wavelength lattice axis~\cite{SupplInf}, see Fig.~\ref{fig:Setup_scheme}(b).}
We operate the magic-wavelength lattice at a depth of $\sim$ 107\,\Erecoil\,and derive atomic temperature via the \newtext{axial} sideband spectra, see Fig.~\ref{fig3:Sisyphus_cooling}(c).
Without Sisyphus cooling, a noticeable red sideband amplitude reflects the atoms' population among the lattice bands with $n_{\mathrm{z}}>0$, as shown by the red curves in Fig.~\ref{fig3:Sisyphus_cooling}(c).
The measured ratio of red and blue sideband areas give a longitudinal temperature
of $\sim$9\,$\mu$K~\cite{Blatt2009_RabiSpectroscopy}.
After applying Sisyphus cooling for 200\,ms, the blue sideband narrows, a signature of lower radial temperature~\cite{Blatt2009_RabiSpectroscopy}. In our 2D Sisyphus cooling, we consistently achieve a radial temperature \newtext{near the} recoil \newtext{limit, 410\,nK.}
We note that the longitudinal temperature (proportional to the red sideband amplitude) also decreases to $\sim\,$0.8\,$\mu$K, despite only applying Sisyphus cooling along the radial directions. Longitudinal cooling occurs due to a motional sideband cooling process: in some regions of the 1388\newtext{-}nm Sisyphus \newtext{potential}, the energy of the $\ket{{^3\mathrm{P}_0}, n_z=n-1}$ state is ac-Stark shifted into resonance with the unperturbed transition, satisfying the longitudinal red sideband cooling condition~\cite{Brown2017_Hyperpolarizability}, as depicted in Fig.~\ref{fig3:Sisyphus_cooling}(d). \newtext{Under a range of experimentally-accessible magic-wavelength trap depths, we find that the red sideband cooling conditions are satisfied. For example, with the same Sisyphus potential used in the $\sim$ 107\,\Erecoil\,magic-wavelength lattice, we also observed a similar longitudinal temperature of $\sim\,$0.8\,$\mu$K even when we intentionally reduced the magic-wavelength lattice depth by half to $\sim$ 55\,\Erecoil.}

\begin{figure}[tb]
\includegraphics[width=0.98\columnwidth]{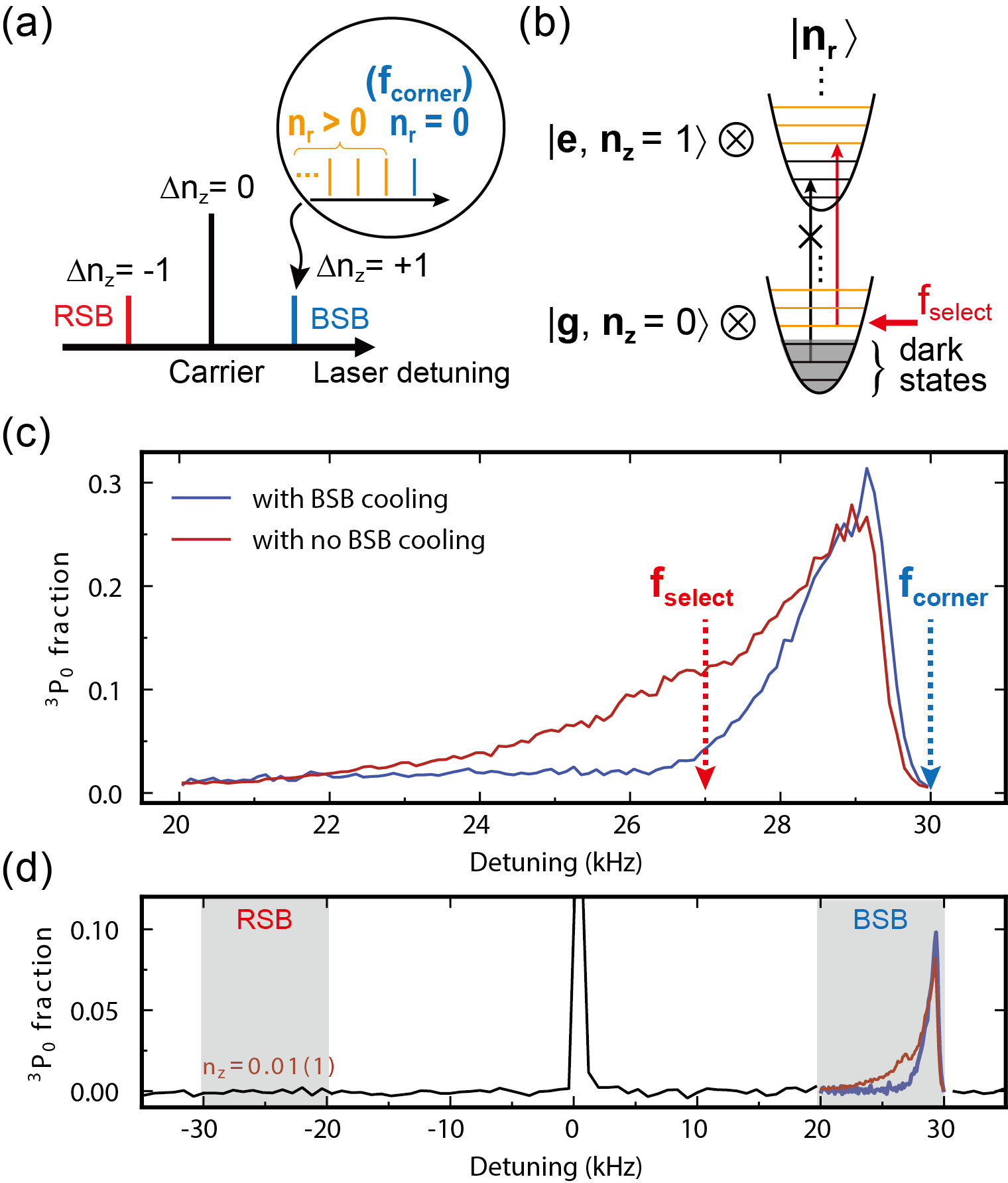}
\caption{\label{fig:Trap_assisted_Sisyphus_subrecoil_cooling} \textbf{Blue sideband assisted velocity selection towards sub-recoil temperature.} (a) Lamb-Dicke regime spectroscopy in a 1D optical lattice: the most intense line represents the carrier, where the red (blue) line represents the 1\textsuperscript{st} order RSB (BSB) transition. Higher radial energies smear the sidebands to the carrier. The \finalEditnewtext{excited radial modes ($\mathrm{n}_{\mathrm{r}}>$0) depicted in the color orange within the BSB.}
(b) Two states addressed by radiation. Those with insufficient radial energy are unexcited, dark states.   Those with sufficient radial energy are referred to bright states and can be easily excited.
(c) We plot the BSB before and after our BSB cooling to sub-recoil temperatures. The upper edge of the BSB, refered to as $f_{\textrm{corner}}$, marks the minimum radial temperature, while $f_{\textrm{select}}$ sets the threshold for dark states. (d) Longitudinal sideband spectra at 61.3(4)\Erecoil, both before and after the BSB cooling. RSB and BSB transitions are highlighted with shaded areas, respectively. Additional longitudinal cooling was applied via two RSB ARP's in both cases, which demonstrates the low 3D temperatures ($\bar{n}_\textrm{z}=0.01(1)$) that we observed.}
\end{figure}

\textit{Motional state selective excitation on the blue sideband.---}We also explore an alternative method for achieving sub-recoil temperatures using the narrow clock transition to selectively excite atoms on the blue sideband ($\Delta n_z$\,=\,$+$1) (BSB) in a 1D optical lattice.
This approach combines the techniques of light-induced evaporation~\cite{Walraven1993LightinducedEvaporation} and cyclic cooling~\cite{Pritchard1987LaserSpectroscopyVIII, Pritchard1983CyclicCooling}, and utilizes information of the radial energy distribution, $E(n_{\textrm{r}})$, encoded in the BSB spectra~\cite{Beloy2020_LightShift, Blatt2009_RabiSpectroscopy}, see Fig.~\ref{fig:Trap_assisted_Sisyphus_subrecoil_cooling}(a). Our method is similar to a recent experiment of motion-selective coherent population trapping (MSCPT)~\cite{Lee2022MotionSelective, Park2022MotionSelective}, where a higher vibration frequency of the trapped atom is selectively excited by Raman beams. \newtext{Here}, by adjusting the frequency of the clock laser ($f_{\textrm{select}}$) below the lattice trap corner frequency ($f_{\textrm{corner}}$), we selectively excite atoms with higher radial motional energy into the \clockstate\,clock state, while leaving those with lower energy in the \groundstate\,state, effectively dark to the selection light, see Fig.~\ref{fig:Trap_assisted_Sisyphus_subrecoil_cooling}(b).
Radially hot atoms are cooled and quenched from the \clockstate\,clock state using the Sisyphus potential.

A cooling cycle consists of the following four steps.
First, using the narrow clock laser, we excite at $f_{\textrm{select}}$ on the BSB, see Fig.~\ref{fig:Trap_assisted_Sisyphus_subrecoil_cooling}(c). 
Second, we apply Sisyphus cooling for 1\,ms with a 1388\newtext{-}nm laser detuning 
$\Delta \approx+50 $\,MHz and an optical power of $\approx100\,\mu\mathrm{W}$~\cite{SupplInf}. Third, using the narrow clock laser, we apply a 2 ms $^{1}\textrm{S}_{0}\,\rightarrow\,^{3}\textrm{P}_0$ adiabatic rapid passage (ARP) on the red sideband ($\Delta n_z$\,=\,$-$1)~\cite{Siegel2024Delocalization}. 
Fourth, we use the Sisyphus cooling again to restore population to $^{1}\textrm{S}_{0}$. 
Starting with a recoil temperature limited sample, we achieved sub-recoil temperatures in 10 cycles, which lasts for a total of 81\,ms.
The radial temperature dropped from 451(35)\,nK to 314(37)\,nK, and reduced the amplitude of the BSB at frequencies below $f_{\textrm{select}}$, see Fig.~\ref{fig:Trap_assisted_Sisyphus_subrecoil_cooling}(c).
However, the atom loss was comparable to an energy filtering method (adiabatically ramping the magic wavelength trap depth down and back up).
Despite the BSB excitation, no noticeable longitudinal heating was measured, see Fig.~\ref{fig:Trap_assisted_Sisyphus_subrecoil_cooling}(d); thus, we expect that atom loss was due to unwanted optical pumping to the $^{3}P_2$ metastable state.
In principle, this loss can be eliminated by adding a repump laser.
We note that the longitudinal cooling of the Sisyphus potential makes ``continuous" BSB cooling possible.
During continuous operation of the clock laser at $f_{\textrm{select}}$ and the Sisyphus potential, we measure subrecoil radial temperatures and $\bar{n}_z$ below 0.32(5).

We demonstrate Sisyphus cooling of $^{171}$Yb atoms with a spatially varying ac-Stark shift on the \clockstate state from 1388\newtext{-}nm laser light, which is also often used for excited clock state depumping.
We enhance the loading of atoms into the magic-wavelength optical lattice and reach subrecoil temperatures with both pulsed and continuous cooling. We show 1D Sisyphus cooling to a temperature of 165(5)\,nK. We create a 2D Sisyphus potential in the transverse plane of the magic-wavelength optical lattice, enabling effective 3D cooling. This approach represents a straightforward modification to the typical 1D optical lattice clock architecture and could prove useful to high performance portable clock systems. We further realize subrecoil temperatures assisted by the ``energy selective excitation" on the blue sideband that is broadened by \finalEditnewtext{the radial motion of the atoms}.

Our cooling does not require dynamically reducing the trapping potentials. Therefore the trap frequency remained high during the cooling process, which could allow for accelerating (runaway) evaporative cooling~\cite{Hung2008PRA_RunawayEvaporative}.
In addition, this method does not require a favorable elastic scattering rate, which is essential for evaporative cooling. This makes it suitable for cooling \finalEditnewtext{fermions}, such as those used in optical lattice clocks.
We expect these techniques to find applicability in other alkaline-earth(-like) atoms~\cite{Katori2008_MercuryLattticeclock, Katori2019_CdMagiceWavelength}. While narrow-line cooling to a few $\mu$K has been recently demonstrated in Cd~\cite{Katori2019_CdMagiceWavelength}, laser cooling in Hg and Mg are currently limited to temperatures of several tens of $\mu$K, preventing efficient lattice trap loading. Meanwhile, lattice depths at their respective magic-wavelengths are also limited by available laser power, making it more challenging to achieve a deep lattice to facilitate loading without a build-up cavity. Through dressing the long-lived state with a periodic ac-Stark shift\textcolor{black}{~\cite{SupplInf, Repumpfootnote}} (Hg: $\mathrm{6s6p }^{3}\mathrm{P}_{0}\rightarrow  \mathrm{6s7s }^{3}\mathrm{S}_{1}$ at 405\,nm~\cite{Katori2008_MercuryLattticeclock}, \newtext{Mg: $\mathrm{3s3p }^{3}\mathrm{P}_{1}\rightarrow  \mathrm{3s4s }^{1}\mathrm{S}_{0}$ at 462\,nm~\cite{Rasel2007_quenchcoolingMg}}), the Sisyphus cooling method could produce deeper cooling, increasing the loading efficiency of shallow UV lattices, as compared to the cooling using the intercombination line.

Here we have demonstrated efficient cooling without requiring high-power clock laser beams or time-varying optical or magnetic fields~\cite{Ludlow2022PRL_YbSubrecoil}. This may enable continuous quantum sensors to operate with high bandwidth, high signal-to-noise ratio, while free from aliasing~\cite{Black2022PRApplied_CWinterferometer, Ludlow2019PRL_CaRamseyBorde, Katori2021RamseyCWclock, Takeuchi_2023_CWSrBeam}.
By controlling the illumination region of the clock light, our method in principle allows for site selective cooling and imaging~\cite{Covey2019PRL_Sr2000timesImaging}. It could also serve as a unique quantum engineering tool for the realization of novel non-equilibrium states~\cite{Christie2018_LatticeQuantumEngineering}, and enabling subsystem readout during a quantum process, such as mid-circuit measurements.

These Sisyphus cooling techniques may be applicable to (anti)hydrogen experiments~\cite{Baker2021LaserCoolingAntiH, Yost2022arxiv_HydrogenDeceleration}. Currently, laser cooling using pulsed narrow-linewidth Lyman-$\alpha$ transition light was recently demonstrated in cooling magnetically trapped anti-hydrogen~\cite{Baker2021LaserCoolingAntiH}. Additionally, there has been a successful demonstration of hydrogen beam deceleration~\cite{Yost2022arxiv_HydrogenDeceleration}. Our work has the potential to contribute to the exploration of anti-hydrogen cooling schemes exploiting the dressed metastable state~\cite{Wu2011_SisyphusCoolHydrogen}, for improved spectroscopic precision. 

We acknowledge useful conversations with Tobias Bothwell.
We thank \textcolor{black}{Vladi Gerginov
 and Wesley Brand} for careful reading of the manuscript and providing insightful comments. K.G. acknowledges the financial support from the \newtext{from the U.S. National Science Foundation under award No.\,2012117}. This work was supported by NIST, ONR, and NSF QLCI Award No.\,2016244.


\begin{thebibliography}{46}
\bibitem [{Sup()}]{Repumpfootnote}%
  \BibitemOpen
  \href@noop {} {\bibinfo  {journal} {\textcolor{black}{In a clock-line mediated Sisyphus cooling scheme employing the ${^3\mathrm{D}_1}$ as the higher-lying excited state, if no repumper is applied, atoms fall out of the cooling cycle once they decay to the ${^3\mathrm{P}_2}$ dark state (branching ratio $\sim$ 1\% for ${^3\mathrm{D}_1}\rightarrow {^3\mathrm{P}_2}$).
However, only a few cooling cycles are required to transition from the initial temperatures of ${^3\mathrm{P}_1}$ Doppler cooling down to the recoil limit}}\ }\BibitemShut {NoStop}%
\end{thebibliography}

%

\beginsupplement

\section*{Supplemental Material}

\section*{Selection of the higher lying excited state for inducing the Sisyphus lattice}
Compared to s-orbitals and other higher lying d-orbitals, we choose the $6s5d\,^{3}\mathrm{D}_{1}$ state as our ac-Stark shift dressing state for two reasons. First, d orbitals have more favorable spontaneous emission decay branching ratios to the $^3\mathrm{P}$ manifold, $^{3}\mathrm{D}_{1}\rightarrow({^3\mathrm{P}_0}, {^3\mathrm{P}_1}, {^3\mathrm{P}_2})$ = (65\%, 34\%, $<$1\%), therefore, atoms preferentially stay within the Sisyphus cooling cycles~\cite{Kyle2012_3D1Lifetime} after scattering multiple \IRTransition~photons. 
For experimental simplicity, we did not apply ${^3\mathrm{P}_2} \rightarrow {^3\mathrm{D}_1}$ repumping during the Sisyphus cooling process. We did not observe density dependent \clockstate ~state two-body loss either.  Consequently, atom loss is currently dominated by undesired decay into the $^{3}\mathrm{P}_{2}$ state. In the future, we anticipate that adding a repumping laser could help preserve all atoms during the Sisyphus cooling. Second, scattering of an infrared photon leads to very low recoil heating of only a few tens of nK~\cite{Hobson2020PRA_SrIRMOT, KatoriPRA2021_IRcooling}.

\begin{figure}[b!]
\includegraphics[width=0.98\columnwidth]{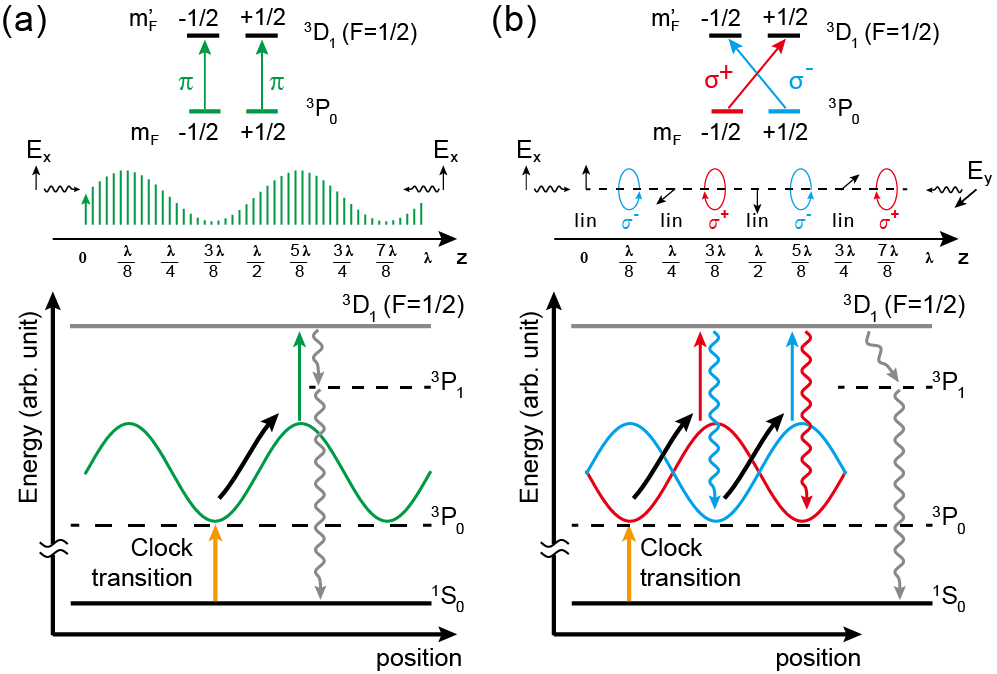}
\caption{\label{fig2:Sisyphus_lattice_scenerio} \textbf{The two types of clock-state light shift gradient in a 1-D Sisyphus cooling in $^{171}$Yb. \newtext{For simplicity, we highlight here coupling to the $^{3}\mathrm{D}_{1}$ $F=1/2$ hyperfine state, though both the $F=1/2,3/2$ states can accommodate Sisyphus potentials.}} (a) \textbf{lin$\parallel$lin configuration:} two counterpropagating co-linearly polarized 1388-nm light address the \IRTransition transition with \newtext{linear} polarization and form an intensity lattice.
Two $^{3}\mathrm{P}_{0}$ clock state sublevels experience the same light shift in the 1388-nm light intensity lattice and form spatially varying dressed state energy levels for the Sisyphus cooling. (b) \textbf{lin$\perp$lin configuration:} two counter-propagating orthogonally linearly polarized beams form a polarization gradient lattice. The two clock state sublevels experience spatially varying light shifts that are displaced by $\lambda/4$. Spontaneous decay from $^{3}\mathrm{D}_{1}$ to the $^{3}\mathrm{P}_{1}$ and $^{3}\mathrm{P}_{0}$ states \newtext{can} both realize the Sisyphus cooling.}
\end{figure}

\section*{Sisyphus lattice: Lin~$\parallel$~Lin v.s. Lin~$\perp$~Lin}
\label{Sec:Sisyphus_cases}
\color{black}

We identify two types of periodic light shift cases in the Sisyphus cooling scheme in $^{171}$Yb ($\mathrm{I}=\frac{1}{2}$), see Fig.~\ref{fig2:Sisyphus_lattice_scenerio}.
Here we only describe cooling in 1D to simplify the discussion, though 2D cooling proceeds along similar lines. 
When counter-propagating 1388-nm beams have parallel-aligned polarizations, it results in an intensity standing wave, see Fig.~\ref{fig2:Sisyphus_lattice_scenerio}(a). 

Atoms experience an intensity lattice from the 1388-nm laser, and the two m$_{\mathrm{F}}$ sublevels of the $^{3}\mathrm{P}_{0}$ clock state experience the same spatially varying light shift.
Once the atoms are excited to the $^{3}\mathrm{P}_{0}$ clock state, they explore the spatial varying dressed clock state energy level before scattering a 1388-nm photon and getting excited to the $^{3}\mathrm{D}_{1}$ state. Atoms have a higher chance of scattering a 1388-nm photon at a position corresponding to the intensity lattice antinode.
Atoms successfully dissipate their energy via cascaded spontaneous emission to $^{1}\mathrm{S}_{0}$, via the $^{3}\mathrm{P}_{1}$ state. 

Alternatively, when the counter-propagating beams have orthogonal linear polarizations(lin$\perp$lin), the intensity is constant along the beam, and atoms instead experience a polarization gradient lattice, as shown in Fig.~\ref{fig2:Sisyphus_lattice_scenerio}(b).  In this case, the Sisyphus potential is mediated through vector polarizability of $^{3}\mathrm{P}_{0}$ from near resonant coupling to a $^{3}\mathrm{D}_{1}$ hyperfine state.
Because of optical pumping among $^{3}\mathrm{P}_{0}$ sublevels, emitted fluorescence photon can be blue-shifted by an amount corresponding to the light shift splitting between the two $^{3}\mathrm{P}_{0}$ clock state sublevels, leading to an extra Sisyphus cooling effect apart from the normal Sisyphus cooling spontaneous decay path to the ground state via $^{3}\mathrm{P}_{1}$.

\section*{Simple Model for ac-Stark Shift}
\mbox{~}

We determine the spatially dependent ac-Stark shift of the $^{3}\mathrm{P}_{0}$ dressed state, in one dimension, by diagonalizing the ac-Stark Hamiltonian
\begin{equation}
    H = H_{a} + H_{LA},
\end{equation}
where $H_{a}$ is the bare atom Hamiltonian and $H_{LA}$ is the atom-field interaction Hamiltonian.
The $H_{a}$ and $H_{LA}$ Hamiltonian's are given by 
\begin{equation}
\label{eqn:Htot}
\begin{aligned}
    H_{a} & = \hbar \sum_{m}  \omega_c \ket{^{3}\mathrm{P}_{0},m}\bra{^{3}\mathrm{P}_{0},m} 
    \\ & + \left (\omega_c+\omega_s\right) \ket{^{3}\mathrm{D}_{1},m}\bra{^{3}\mathrm{D}_{1},m},
\end{aligned}
\end{equation}
\begin{equation}
\begin{aligned}
 H_{LA}   & = \frac{\hbar\Omega_S(z)}{2} \sum_{m,m',\epsilon} \left[ p_{\epsilon}(z)  \delta_{m,m'+\epsilon} C_{m,\epsilon,m'} \right]   \\ & \times e^{-i(\omega_S+\delta_S )t} \ket{{}^3P_0,m}\bra{{}^3D_1,m'} + h.c.,
\end{aligned}
\end{equation}
where $\Omega_{S}(z)$ is the Rabi frequency for the 1388\,nm Sisyphus cooling laser, $\delta_S$ is its detuning, $\omega_{c}$ is the ${^1\mathrm{S}_0} - {^3\mathrm{P}_0}$ transition frequency, $\omega_s$ is the ${^3\mathrm{P}_0} - {^3\mathrm{D}_1}$ transition frequency, $p_{\epsilon}(z)$ is the coefficient of the normalized polarization vector at a location $z$, $\epsilon =\pm1,0$ is the polarization, $C_{m,\epsilon,m'}$ is the Clebsh-Gordan coefficient, and $m$ ($m'$) is the projection of the initial (final) state's spin onto the $z$ axis. 
We choose the quantization axis, $\hat{z}$, to be oriented along the $\vec{k}$ vector of the 1388-nm light.

\begin{figure}[tb]
\includegraphics[width=0.98\columnwidth]{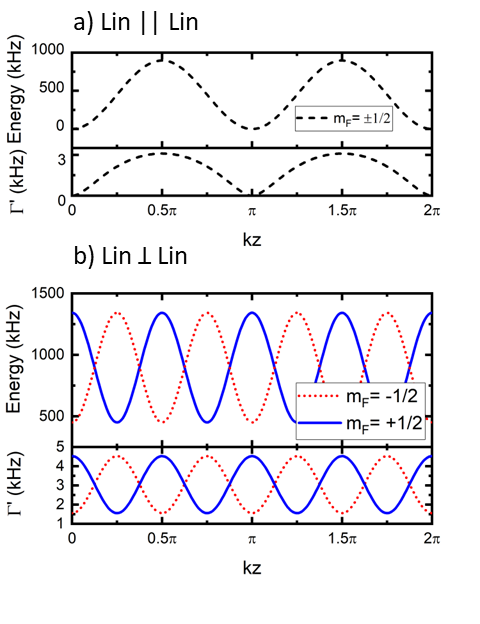}
\caption{\label{fig:dressedstates} \textbf{Simulation of the Sisyphus potential and 1388\,nm scattering rate.} 
We calculate for a 1D Sisyphus potential with 1388\,nm Sisyphus cooling laser detuning, $\delta_S= +150\,\mathrm{MHz}$, and laser power,   $P_\mathrm{1388\,nm}=80\,\mu\mathrm{W}$.
The Sisyphus lattice height and corresponding scattering rate $\Gamma^{'}$ is plotted for the configurations of Lin$\parallel$Lin (a) and Lin$\perp$Lin (b).
}
\end{figure}

The eigenvalues of $H$ yield the spatially dependent ac-Stark shift, and are plotted in the rotating frame for two polarization cases in Fig.~\ref{fig:dressedstates} for a 1D Sisyphus potential.
Lin $\parallel$ Lin polarization, has $p_0= 0$ and $p_{+1}=p_{-1}=1/\sqrt{2}$, while Lin $\perp$ Lin polarization has $p_\pi= 0$, $p_{+1} = \sin^2(2kz)$, $p_{-1}= \cos^2(2kz)$.
The scattering rate of $\ket{^{3}\mathrm{P}_{0},m}$, $\Gamma^{'}$, is calculated from the natural linewidth of the 1388-nm transition times the population in $\ket{^{3}\mathrm{D}_{1},m}$.
$\Gamma^{'}$ is seen to be maximum at locations of maximum potential energy. 

We compare this theory to an experimental measure of the Sisyphus lattice height.
The ${}^3\mathrm{P}_0$ state ac-Stark shift is measured via spectroscopy on the 578nm clock transition.
We measure the frequency shift of the ${}^3\mathrm{P}_0$ resonance as a function of intensity for a single 1388nm running wave at $\delta_S=-50(10)$ MHz. 
Extrapolated to our estimates of the peak intensities used for free space cooling (shown in Table \ref{tab:typicalvalues}), we find a shift of $\sim 0.5-1$ MHz. 
We compare this result to estimates from the simple model above, and observe reasonable agreement within a factor of $\sim 2-4$.

\section*{Typical Parameters Used During Cooling}
\begin{table}[h]
    \centering
    \begin{tabular*}{\columnwidth}{@{\extracolsep{\fill}}lcccc}
      Experiment &  \makecell{$\delta_{s}$ \\ (MHz)} & \makecell{$P_{s}$ x/y \\($\mu\mathrm{W}$)} & \makecell{Theory:\\ $U_{\mathrm{Sisyphus}}$\\ (kHz)} & \makecell{$\Omega_{\mathrm{clk}}$ \\(kHz)}\\
      \hline\noalign{\smallskip} \hline \noalign{\smallskip}
       Free Space   & 50 & 80/110 & \num{1e3} & $\sim$5\\
       1D Cooling & 150 & 0/130 & \num{5e2} & \makecell{0.760(14) /\\ 2.799(41)}\\ 
       3D Cooling & 150 & 130/130 & \num{2e3} & 2.7 \\
      BSB Excitation & 50 & 85/150 & \num{2e3} & 1.6

    \end{tabular*}
    \caption{\textbf{Typical parameters used in different cooling outlined in the main text.}  The typically used Sisyphus laser power ($P_{s}$) is shown for each counter-propagating beam for the two orthogonal directions x and y. \finalEditnewtext{We note that there is an unequal beam intensity between x and y due to the window transmission loss in our setup. We observed our result is insensitive to this imbalance. Here, the theoretical Sisyphus lattice depth takes into account the window transmission loss, $T_x =\,\sim 40\%$ and $T_y =\,\sim 60\%$.}
    }
    \label{tab:typicalvalues}
\end{table}

\end{document}